\documentclass[a4paper,runningheads]{llncs}

\usepackage{amssymb}
\usepackage{graphicx}
\usepackage{makeidx}
\usepackage{hyperref}

\usepackage{url}
\urldef{\mailsa}\path|Krzysztof.szajowski@pwr.wroc.pl|
\newcommand{\keywords}[1]{\par\addvspace\baselineskip
\noindent\keywordname\enspace\ignorespaces#1}

\newtheorem{ass}[definition]{Assumptions}

\def\bbE{{\mathbb E}}
    \def\BbbE{\mathbb E}
    \def\BbbN{\mathbb N}

\def\bP{{\bf P}}
\def\bE{{\bf E}}

\def\cC{{\mathcal C}} %albo \cal
\def\cF{{\mathcal F}}
\def\cB{{\mathcal B}}
\def\cT{{\mathcal T}}
\def\cW{{\mathcal W}}
\def\cL{{\mathcal L}}

\def\frt{{\mathfrak t}}
\def\frF{{\mathfrak F}}
\def\frS{{\mathfrak S}}
\def\frN{{\mathfrak N}}

\def\vecPi{\overrightarrow{\Pi}}

\def\one{{\mathbb I}}
       \def\vecxi{\overrightarrow{\xi}}
       \def\vecx{\overrightarrow{x}}
       \def\vecX{\overrightarrow{X}}

\begin{document}

\mainmatter  % start of an individual contribution

\title{Multi-variate Quickest Detection of Significant Change Process}

\titlerunning{Multivariate Quickest Detection}

\author{Krzysztof Szajowski%
%\thanks{Please note that the LNCS Editorial assumes that all authors have used the western naming convention, with given names preceding surnames. This determines the structure of the names in the running heads and the author index.}%
%\and Ursula Barth
}
\authorrunning{K.Szajowski}

\institute{Institute of Mathematics and Computer Science, Wroc\l{}aw University of Technology,\\
 Wybrze\.ze Wyspia\'{n}skiego 27, 50-370 Wroc\l{}aw, Poland\\
\mailsa\\
\url{http://www.im.pwr.wroc.pl/\~szajow}}


\toctitle{Multivariate Quickest Detection}
\tocauthor{K.Szajowski}
\maketitle

\begin{abstract}
The paper deals with a mathematical model of a surveillance system based on a net of sensors. The signals acquired by each node of the net are Markovian process, have two different transition probabilities, which depends on  the presence or absence of a intruder nearby. The detection of the transition probability change at one node should be confirmed by a detection of similar change at some other sensors. Based on a simple game the model of a fusion center is then constructed. The aggregate function defined on the net is the background of the definition of a non-cooperative stopping game which is a model of the multivariate disorder detection. 

\end{abstract}

\section{\label{intro} Introduction}
The aim of this consideration is to construct the mathematical model of a multivariate surveillance system. It is assumed that there is net $\frN$ of $p$ nodes. At each node the state is the signal at moment $n\in \BbbN$ which is at least one coordinate of the vector $\vecx_n\in\bbE\subset\Re^m$. The distribution of the signal at each node has two forms and depends on \emph{a pure} or \emph{a dirty} environment of the node. The state of the system change dynamically. We consider the discrete time observed signal as $m\geq p$ dimensional process defined on the fixed probability space $(\Omega,\cF,\bP)$. The observed at each node process is Markovian with two different transition probabilities (see \cite{sarsza11:transition} for details). In the signal the visual consequence of the transition distribution changes at moment $\theta_i$, $i\in\frN$ is a change of its character. To avoid false alarm the confirmation from other nodes is needed. The family of subsets (coalitions) of nodes are defined in such a way that the decision of all member of some coalition is equivalent with the claim of the net that the disorder appeared. It is not sure that the disorder has had place. The aim is to define the rules of nodes and a construction of the net decision based on individual nodes claims. Various approaches can be found in the recent research for description or modeling of such systems (see e.g. \cite{tarvee08:sensor}, \cite{ragvee10:Markov}). The problem is quite similar to a pattern recognition with multiple algorithm when the fusions of individual algorithms results are unified to a final decision. The proposed solution will be based on a simple game and the stopping game defined by a simple game on the observed signals. It gives a centralized, Bayesian version of the multivariate detection with a common fusion center that it has perfect information about observations and \emph{a priori} knowledge of the statistics about the possible distribution changes at each node. Each sensor (player) will declare to stop when it detects disorder at his region. Based on the simple game the sensors' decisions are aggregated to formulate the decision of the fusion center. The  sensors' strategies are constructed as an equilibrium strategy in a non-cooperative game with a logical function defined by a simple game (which aggregates their decision). 

The general description of such multivariate stopping games has been formulated by Kurano, Yasuda and Nakagami in the case when the aggregation function is defined by the voting majority rule \cite{kyn} or the monotone voting strategy \cite{ynk} and the observed sequences of the random variables are independent, identically distributed. It was Ferguson~\cite{fer05:MR2104375} who substituted the voting aggregation rules by a simple game. The Markov sequences have been investigated by the author and Yasuda~\cite{szayas95:voting}.

The model of detection the disorder at each sensor are presented in the next section. It allows to define the individual payoffs of the players (sensors). Section~\ref{coopnobcoop} introduces the aggregation method based on a simple game of the sensors. Section~\ref{noncooperative} contains derivation of the non-cooperative game and existence theorem for equilibrium strategy. The final decision based on the state of the sensors is given by the fusion center and it is described in Section~\ref{strategiesOFsensors}. The natural direction of further research is formulated also in the same section. A conclusion and resume of an algorithm for rational construction of the surveillance system is included in  Section~\ref{finalremarks}.   

\section{\label{disorderONsensor}Detection of disorder at sensors}
Following the consideration of Section~\ref{intro}, let us suppose that the process $\{\vecX_n,n\in\BbbN\}$, $\BbbN=\{0,1,2,\ldots\}$, is observed sequentially in such a way that each sensor, \emph{e.g.} $r$ (gets its coordinates in the vector $\vecX_n$ at moment $n$). By assumption, it is a stochastic sequence that has the Markovian structure given random moment $\theta_r$, in such a way that the process after $\theta_r$ starts from state $\vecX_{n\; \theta_r-1} $. The objective is to detect these moments based on the observation of $\vecX_{n\; \cdot}$ at each sensor separately. There are some results on the discrete time case of such disorder detection which generalize the basic problem stated by Shiryaev~in~\cite{shi61:detection} (see e.g. Brodsky and Darkhovsky~\cite{brodar93:nonparametr}, Bojdecki~\cite{boj79:disorder}, Yoshida~\cite{yos83:complicated}, Szajowski~\cite{sza92:detection}) in various directions.

Application of the model for the detection of traffic anomalies in networks has been discussed by Tartakovsky et al.~\cite{tarroz06:intrusions}. The version of the problem when the moment of disorder is detected with given precision will be used here (see~\cite{sarsza11:transition}).

\subsection{\label{sformProblem} Formulation of the problem}
%\vspace{-.5cm}
The observable random variables $\{\vecX_n\}_{n \in \BbbN}$ are consistent with the filtration $\mathcal{F}_n$ (or $\cF_n = \sigma(\vecX_0,\vecX_1,\ldots,\vecX_n)$). The random vectors $\vecX_n$ take values in $(\bbE, \mathcal{B})$, where $\bbE\subset\Re^m$. On the same probability space there are defined unobservable (hence not measurable with respect to $\cF_n$) random variables $\{\theta_r\}_{r=1}^m$ which have the geometric distributions:
%\setlength\arraycolsep{2pt}
%\vspace{-3.67ex}
%\small
\begin{eqnarray}
\label{rozkladyTeta}
\bP(\theta_r = j) = p_r^{j-1}q_r, \mbox{ $q_r=1-p_r \in (0,1)$, $j=1,2,\ldots$.}
\end{eqnarray}

The sensor $r$ follows the process which is based on switching between two, time homogeneous and independent, Markov processes $\{X_{rn}^i\}_{n \in \BbbN}$, $i=0,1$, $r\in\frN$ with the state space $(\bbE, \mathcal{B})$, both independent of $\{\theta_r\}_{r=1}^m$. Moreover, it is assumed that the processes $\{X_{rn}^i\}_{n \in \BbbN}$ have transition densities with respect to the $\sigma$-finite measure $\mu$, i.e., for any $B\in\cB$ we have  
\begin{eqnarray}\label{TransProbab}
\bP_x^{i}(X_{r1}^{i}\in B)&=&\bP(X_{r1}^{i}\in B|X_{r0}^{i}=x)=\int_Bf_x^{ri}(y)\mu(dy).
\end{eqnarray}
The random processes $\{X_{rn}\}$, $\{X_{rn}^0\}$, $\{X_{rn}^1\}$ and the random variables $\theta_r$ are connected via the rule: conditionally on $\theta_r = k$
\begin{eqnarray*}
X_{rn}&=&\left\{\begin{array}{ll}
X_{rn}^0,&\mbox{ if $k>n$,}\\
X_{r\;n+1-k}^1,&\mbox{ if $k\leq n$,}
\end{array}
\right.
\end{eqnarray*}
where $\{X_{rn}^1\}$ is started from $X_{r\;k-1}^0$ (but is otherwise independent of $X_{r\;\cdot}^0$).

For any fixed $d \in \{0,1,2,\ldots\}$ we are looking for the stopping time $\tau_r^{*}\in \cT$ such that
\begin{equation}
\label{PojRozregCiagowMark-Problem}
  \bP_x( | \theta_r - \tau_r^{*} | \leq d ) = \sup_{\tau \in \frS^X} \bP_x( | \theta_r - \tau | \leq d )
\end{equation}
where $\frS^X$ denotes the set of all stopping times with respect to the filtration
$\{\mathcal{F}_n\}_{n \in \BbbN}$. The parameter $d$ determines the precision level of detection and it can be different for too early and too late detection.

\subsection{Construction of the optimal detection strategy}
In \cite{sarsza11:transition} the construction of $\tau^{*}$ by transformation of the problem to the optimal stopping problem for the Markov process $\vecxi$ has been made, such that   $\vecxi_{rn}=(\underline{\vecX}_{r\;n-1-d,n},\Pi_n$), where  $\underline{\vecX}_{r\;n-1-d,n}=(\vecX_{r\;n-1-d},\ldots,\vecX_{r\;n})$ and $\Pi_{rn}$ is the posterior process:
\begin{eqnarray}
    \Pi_{r0} &=& 0,\nonumber\\
    \Pi_{rn} &=& \bP_x\left(\theta_r \leq n \mid \mathcal{F}_n\right),\; n = 1, 2, \ldots  \nonumber
\end{eqnarray}
which is designed as information about the distribution of the disorder instant $\theta_r$. In this equivalent the problem of the payoff function for sensor $r$ is $h_r(\vecx_{r\;d+2},\alpha)$.

\section{\label{coopnobcoop} The aggregated decision via the cooperative game}% Cooperative and non-cooperative games of the sensors}
There are various methods combining the decisions of several classifiers or sensors. 
Each ensemble member contributes to some degree to the decision at any point of the sequentially delivered states. The fusion algorithm takes into account all the decision outputs from each ensemble member and comes up with an ensemble decision. When classifier outputs are binary, the fusion algorithms include the majority voting \cite{lamkrz94:majority}, \cite{LamSue97:pattern}, na\"{\i}ve Bayes combination~\cite{dompaz93:bayesian}, behavior knowledge space~\cite{huasue95:multiple}, probability approximation~\cite{kankimkim97:optimal} and singular value decomposition~\cite{mer99:correspondence}.

The majority vote is the simplest. The extension of this method is a simple game.        

\subsection{A simple game}% and the monotone stopping rule}
Let us assume that there are many nodes absorbing information and make decision if the disorder has appeared or not. The final decision is made in the fusion center which aggregates information from all sensors. The nature of the system and their role is to detect intrusion in the system as soon as possible but without false alarm. 

The voting decision is made according to the rules of \emph{a simple game}. Let us recall that a coalition is a subset of the players.  Let ${\cC}=\{C:C\subset \frN\}$ denote the class of all coalitions.  
\begin{definition}(see \cite{Owe95:MR1355082}, \cite{fer05:MR2104375})
\emph{A simple game} is coalition game having the characteristic function, $\phi(\cdot):\cC\rightarrow\{0,1\}$.      
\end{definition}
Let us denote $\cW=\{C\subset\frN:\phi(C)=1\}$ and ${\cL}=\{C\subset\frN:\phi(C)=0\}$. The coalitions in $\cW$ are called the winning coalitions, and those from $\cL$ are called the losing coalitions.
\begin{ass}
By assumption the characteristic function satisfies the properties:
\begin{enumerate}
\item\label{equat1} $\frN\in\cW$;
\item\label{equat2} $\emptyset\in \cL$;
\item\label{equat3} (the monotonicity): $T\subset S\in \cL$ implies $T\in \cL$.
\end{enumerate}
\end{ass}

\subsection{The aggregated decision rule}
When the simple game is defined and the players can vote presence or absence, $x_i=1$ or $x_i=0$, $i\in\frN$, of the intruder then the aggregated decision is given by the logical function
\begin{equation}\label{aggregateFUNCTION}
\pi(x_1,x_2,\ldots,x_p)=\sum_{C\in\cW}\prod_{i\in C}x_i\prod_{i\notin C}(1-x_i).
\end{equation}
For the logical function $\pi $ we have (cf \cite{ynk}) 
\[
\pi (x^1,\ldots ,x^p)=
x^i\cdot \pi (x^1,\ldots ,\stackrel{i}{\breve{1}}% 
,\ldots,x^p)
+\overline{x}^i\cdot \pi (x^1,\ldots ,\stackrel{i}{\breve{0}}%
,\ldots ,x^p).
\]
%%%TUTAJ
%\section{Formulation of the problem\label{formulation}} 
\section{\label{noncooperative}A non-cooperative stopping game}
Following the results of the author and Yasuda~\cite{szayas95:voting} the multilateral stopping of a Markov chain problem can be described in the terms of the notation used in the non-cooperative game theory (see \cite{nas51:noncoop}, \cite{dresh81:games}, \cite{moulin}, \cite{Owe95:MR1355082}). 
Let $(\vecX_n,\frF_n,{\bP}_x)$, $n=0,1,2,\ldots ,N$, be a homogeneous Markov chain with state space $(\BbbE,\cB)$. The horizon can be finite or infinite. The players are able to observe the Markov chain sequentially. Each player has their utility function $f_i: \BbbE\rightarrow \Re $, $i=1,2,\ldots ,p$, such that ${\bE}_x|f_i(\vecX_1)|<\infty $. If process is not stopped at moment $n$, then each player, based on $\frF_n,$ can declare independently their willingness to stop the observation of the process.

\begin{definition}
{\rm (see \cite{ynk})} An individual stopping strategy of the player $i$ (ISS) is the sequence of random variables $\{\sigma _n^i\}_{n=1}^N$, where $\sigma_n^i:\Omega \rightarrow \{0,1\}$, such that $\sigma _n^i$ is $\frF_n$-measurable.
\end{definition}

The interpretation of the strategy is following. If $\sigma _n^i=1$ then player $i$ declares that they would like to stop the process and  accept the realization of $X_n$. Denote $\sigma ^i=(\sigma _1^i,\sigma _2^i,\ldots ,\sigma _N^i)$ and let $\frS^i$ be the set of ISSs of player $i$, $i=1,2,\ldots ,p$. Define 
\[
\frS=\frS^1\times \frS^2\times \ldots\times \frS^p. 
\]
The element $\sigma =(\sigma ^1,\sigma ^2,\ldots ,\sigma ^p)^T\in \frS$ will be called the stopping strategy (SS). The stopping strategy $\sigma \in \frS$ is a random matrix. The rows of the matrix are the ISSs. The columns are the decisions of the players at successive moments. The factual stopping of the observation process, and the players realization of the payoffs is defined by the stopping strategy exploiting $p$-variate logical function. Let $\pi :\{0,1\}^p\rightarrow \{0,1\}$. In this stopping game model the stopping strategy is the list of declarations of the individual players. The aggregate function $\pi$ converts the declarations to an effective stopping time.

\begin{definition}
A stopping time $\frt_\pi (\sigma )$ generated by the SS $\sigma \in \frS$ and the aggregate function $\pi $ is defined by 
\[
\frt_\pi (\sigma )=\inf \{1\leq n\leq N:\pi (\sigma _n^1,\sigma _n^2,\ldots ,\sigma _n^p)=1\}
\]
$(\inf (\emptyset )=\infty )$. Since $\pi $ is fixed during the analysis we skip index $\pi $ and write $\frt(\sigma )=\frt_\pi (\sigma )$. 
\end{definition}

We have $\{\omega \in \Omega : \frt_\pi (\sigma )=n\} =\bigcap\nolimits_{k=1}^{n-1}\{\omega \in \Omega : \pi (\sigma _k^1,\sigma
_k^2,\ldots,\sigma _k^p)=0\}\cap \{\omega \in \Omega :\pi (\sigma_n^1,\sigma _n^2,\ldots,\sigma _n^p)=1\}\in \frF_n$, then the random  variable $\frt_\pi (\sigma )$ is stopping time with respect to
$\{\frF_n\}_{n=1}^N$. 
For any stopping time $\frt_\pi (\sigma )$ and $i\in \{1,2,\ldots ,p\}$, let
\[
f_i(X_{\frt_\pi (\sigma )})=\left\{
\begin{array}{ll}
f_i(X_n) & \mbox{if }\frt_\pi (\sigma )=n\mbox{,} \\
\limsup_{n\rightarrow \infty }f_i(X_n) & \mbox{if }\frt_\pi (\sigma )=\infty 
\end{array}
\right.
\]
(cf \cite{shi}, \cite{szayas95:voting}). If players use SS $\sigma \in \frS$ and the individual preferences are converted to the effective stopping time by the aggregate rule $\pi $, then player $i$ gets $f_i(X_{\frt_\pi (\sigma )})$. 

Let ${}^{*}\!\sigma =({}^{*}\!\sigma ^1,{}^{*}\!\sigma ^2,\ldots ,{}^{*}\!\sigma ^p)^T$  be fixed SS. Denote 
\[
{}^{*}\!\sigma (i)=({}^{*}\!\sigma ^1,\ldots ,{}^{*}\!\sigma ^{i-1},\sigma ^i,{}^{*}\!\sigma ^{i+1},\ldots,{}^{*}\!\sigma^p)^T.
\]

\begin{definition}
\label{equdef}{\rm (cf. \cite{szayas95:voting})} Let the aggregate rule $\pi $ be fixed. The strategy
${}^{*}\!\sigma =({}^{*}\!\sigma ^1,{}^{*}\!\sigma ^2,\ldots ,{}^{*}\!\sigma ^p)^T\in \frS$ is an equilibrium strategy with respect to $\pi $ if for each $i\in \{1,2,\ldots ,p\}$ and any $\sigma ^i\in \frS^i$ we have
\begin{equation} 
{\bE}_xf_i(\vecX_{\frt_\pi ({}^{*}\!\sigma)})\geq  {\bE}_xf_i(\vecX_{\frt_\pi({}^{*}\!\sigma(i))}). 
\label{defequ}
\end{equation}
\end{definition}
The set of SS $\frS$, the vector of the utility functions $f=(f_1,f_2,\ldots, f_p)$ and the monotone rule $\pi $ define the non-cooperative game $\cal{G}$ = ($\frS$,$f$,$\pi$). The construction of the equilibrium strategy $ {}^{*}\!\sigma \in \frS$ in $\cal{G}$ is provided in \cite{szayas95:voting}. For completeness this construction will be recalled here. Let us define an individual stopping set on the state space. This set describes the ISS of the player. With each ISS of player $i$ the sequence of stopping events $D_n^i=\{\omega :\sigma _n^i=1\}$ combines. For each aggregate rule $\pi$ there exists the corresponding set value function $\Pi :\frF\rightarrow \frF$ such that $\pi (\sigma_n^1,\sigma _n^2,\ldots ,\sigma _n^p)= \pi \{\one_{D_n^1}, \one_{D_n^2},\ldots,\one_{D_n^p}\}= \one_{\Pi(D_n^1,D_n^2,\ldots,D_n^p)}$. For solution of the considered game the important class of ISS and the stopping events can be defined by subsets ${\it{C}}^i \in \cal{B}$ of the state space $\BbbE$. A given set ${\it{C}}^i\in\cal{B}$ will be called the stopping set for player $i$ at moment $n$ if $D_n^i= \{\omega :X_n\in {\it{C}}^i\}$ is the stopping event. 

For the logical function $\pi $ we have  
\[
\pi (x^1,\ldots ,x^p)=
x^i\cdot \pi (x^1,\ldots ,\stackrel{i}{\breve{1}}% 
,\ldots,x^p)
+\overline{x}^i\cdot \pi (x^1,\ldots ,\stackrel{i}{\breve{0}}%
,\ldots ,x^p).
\]
It implies that for $D^i\in \frF$
\begin{equation}
\begin{array}{ll}
\Pi (D^1,\ldots ,D^p)=
& \{D^i\cap \Pi (D^1,\ldots,\stackrel{i}{\breve{\Omega}} , \ldots ,D^p)\} \\
& \cup \{\overline{D}^i\cap \Pi (D^1,\ldots ,\stackrel{i}{\breve{
\emptyset}},\ldots ,D^p)\}.
\end{array}
\label{decomposition}
\end{equation}

Let $f_i$, $g_i$ be the real valued, integrable (i.e. ${\bf E}_x|f_i(X_1)|<\infty $) function defined on $\BbbE$. For fixed $D_n^j$, $ j=1,2,\ldots ,p$, $j\neq i$, and ${\it{C}}^i\in \cal{B}$ define
\[
\psi ({\it{C}}^i)={\bf E}_x\left[f_i(X_1)\one_{{}^i\!D_1(D_1^i)}+ 
%{\bf E}_x
g_i(X_1)\one_{\overline{{}^i\!D_1(D_1^i)}}\right] 
\]
where
${}^i\!D_1(A)=\Pi (D_1^1,\ldots ,D_1^{i-1},A,D_1^{i+1},\ldots ,D_1^p)$ and 
$D_1^i=\{\omega :X_n\in {\it{C}}^i\}$. Let $a^{+}=\max \{0,a\}$ and $a^{-}=\min \{0,-a\}$. 
\begin{lemma} $\label{optimal}$
Let $f_i$, $g_i$, be integrable and let ${\it{C}}^j\in \cal{B}$,
$j=1,2,\ldots,p$, 
$j\neq i$, be fixed. Then the set ${}^{*}\!{\it{C}}^i=\{x\in 
\BbbE:f_i(x)-g_i(x)\geq 0\}\in
\cal{B}$ is such that
\[
\psi ({}^{*}\!{\it{C}}^i)=\sup\limits_{{\it{C}}^i\it{\in }\cal{B} }\psi
({\it{C}}^i)
\]
and
\begin{eqnarray}
\psi ({}^{*}\!{\it{C}}^i)
& = & {\bf E}_x(f_i(X_1)-g_i(X_1))^{+}\one_{{}^{i}\!D_1(\Omega )}
\label{optset} \\
& & - {\bf E}_x(f_i(X_1)-g_i(X_1))^{-}\one_{{}^{i}\!D_1(\Omega )} +{\bf
E}_xg_i(X_1). 
\nonumber
\end{eqnarray}
\end{lemma}

Based on Lemma \ref{optimal} we derive the recursive formulae defining the
equilibrium point and the equilibrium payoff for the finite horizon game. 

\subsection{The finite horizon game\label{finite}} 

Let horizon $N$ be finite. If the equilibrium strategy ${}^{*}\!\sigma $ exists, then we denote $v_{i,N}(x)={\bf E}_xf_i(X_{t({}^{*}\!\sigma )})$ the equilibrium payoff of $i$-th player when $X_0=x$. For the backward induction we introduce a useful notation. Let $\frS_n^i=\{\{\sigma_k^i\},k=n,\ldots ,N\}$ be the set of ISS for moments $n\leq k\leq N$ and $\frS_n=\frS_n^1\times \frS_n^2\times \ldots \times\frS_n^p$. The SS for moments not earlier than $n$ is ${}^n\!\sigma =({}^n\!\sigma ^1,{}^n\!\sigma ^2,\ldots ,{}^n\!\sigma^p) \in \frS_n$, where ${}^n\!\sigma ^i=(\sigma _n^i,\sigma _{n+1}^i,\ldots,\sigma_N^i)$.
Denote
\[
t_n=t_n(\sigma )=t(^n\sigma )=\inf \{n\leq k\leq N:\pi (\sigma _k^1,\sigma
_k^2,\ldots ,
\sigma _k^p)=1\}
\]
to be the stopping time not earlier than $n$. 

\begin{definition}
The stopping strategy ${}^{n*}\!\sigma =({}^{n*}\!\sigma ^1,{}^{n*}\!\sigma^2,\ldots,{}^{n*}\!\sigma ^p)$ is an equilibrium in $\frS_n$ if
\[
\begin{array}{ll}
 {\bf E}_x f_i(X_{t_n({}^{*}\!\sigma   )}) \geq 
 {\bf E}_x f_i(X_{t_n({}^{*}\!\sigma(i))}) & {\bf P}_x-\mbox{a.e.}
\end{array}
\]
for every $i\in \{1,2,\ldots ,p\}$, where \[
{}^{n*}\!\sigma (i)=({}^{n*}\!\sigma^1,\ldots,
{}^{n*}\!\sigma^{i-1},{}^n\sigma^i,
{}^{n*}\!\sigma^{i+1},\ldots,{}^{n*}\!\sigma ^p). \]
\end{definition}

Denote
\[
v_{i,N-n+1}(X_{n-1})={\bf E}_x[f_i(X_{t_n({}^{*}\!\sigma )})|\frF_{n-1}]
= {\bf E}_{X_{n-1}}f_i(X_{t_n({}^{*}\!\sigma )}). \]
At moment $n=N$ the players have to declare to stop and $v_{i,0}(x)=f_i(x)$. Let us assume that the process is not stopped up to moment $n,$ the players are using the equilibrium strategies ${}^{*}\!\sigma _k^i$, $i=1,2,\ldots ,p,$ at moments $k=n+1,\ldots ,N$. Choose player $i$ and assume that other players are using the equilibrium strategies ${}^{*}\!\sigma _n^j$, $j\neq i$, and player $i$ is using strategy $\sigma_n^i$ defined by stopping set ${\it{C}}^i$. Then the expected payoff $\varphi_{N-n}(X_{n-1},{\it{C}}^i) $ of player $i$ in the game starting at moment $n$, when the state of the Markov chain at moment $n-1$ is $X_{n-1\mbox{,}}$ is equal to
\[
 \varphi _{N-n}(X_{n-1},{\it{C}}^i)=
 {\bf E}_{X_{n-1}}\left[f_i(X_n)\one_{{}^{i*}\!D_n(D_n^i)}+ 
 %{\bf E}_x
 v_{i,N-n}(X_n)
 \one_{\overline{{}^{i*}\!D_n(D_n^i)}}\right], 
\]
where
${}^{i*}\!D_n(A)=\Pi({}^{*}\!D_n^1,\ldots,{}^{*}\!D_n^{i-1},A, 
{}^{*}\!D_n^{i+1},\ldots,{}^{*}\!D_n^p)$. 

By Lemma \ref{optimal} the conditional expected gain $\varphi _{N-n}(X_{N-n}, 
{\it{C}}^i)$ attains the maximum on the stopping set
${}^{*}\!{\it{C}}_n^i=\{x\in
\bbE:f_i(x)-v_{i,N-n}(x)\geq 0\}$ and \setcounter{equation}{0}
\begin{equation}
\begin{array}{lll}
v_{i,N-n+1}(X_{n-1})
&=&{\bf E}_x[(f_i(X_n)-v_{i,N-n}(X_n))^{+} 
\one_{{}^{i*}\!D_n(\Omega)}|\frF_{n-1}] \\
& &- {\bf E}_x[(f_i(X_n)-v_{i,N-n}(X_n))^{-} 
\one_{{}^{i*}\!D_n(\emptyset)}|\frF_{n-1}] \\
& &+ {\bf E}_x[v_{i,N-n}(X_n)|\frF_{n-1}] 
\end{array} \label{valueatn}
\end{equation}
${\bf P}_x-$a.e..
It allows to formulate the following construction of the equilibrium strategy and the equilibrium value for the game $\cal{G}$.

\begin{theorem}
In the game $\cal{G}$with finite horizon $N$ we have the following solution. 
\begin{description}
\item[(i)] The equilibrium value $v_i(x)$, $i=1,2,\ldots ,p$, of the game 
$\cal{G}$ can be calculated recursively as follows: \begin{enumerate}
\item $v_{i,0}(x)=f_i(x)$;
\item For $n=1,2,\ldots ,N$ we have ${\bf P}_x-$a.e. 
\small
\begin{eqnarray*}
v_{i,n}(x)={\bf E}_x[(f_i(X_{N-n+1})-v_{i,n-1}(X_{N-n+1}))^{+} 
\one_{{}^{i*}\!D_{N-n+1}(\Omega )}|\frF_{N-n}] \\ 
- {\bf E}_x[(f_i(X_{N-n+1})-v_{i,n-1}(X_{N-n+1}))^{-} 
\one_{{}^{i*}\!D_{N-n+1}(\emptyset )}|\frF_{N-n}] \\ 
+ {\bf E}_x[v_{i,n-1}(X_{N-n+1})|\frF_{N-n}], 
\end{eqnarray*}
for $i=1,2,\ldots ,p$.
\normalsize
\end{enumerate}
\item[(ii)] The equilibrium strategy ${}^{*}\!\sigma \in \frS$ is 
defined by the SS of the players ${}^{*}\!\sigma _n^i$, where
${}^{*}\!\sigma _n^i=1$ if
$ X_n\in {}^{*}\!{\it{C}}_n^i$, and
${}^{*}\!{\it{C}}_n^i=\{x\in \bbE : f_i(x)-v_{i,N-n}(x) \geq 0\}$,
$n=0,1,\ldots ,N$.
\end{description}

We have $v_i(x)=v_{i,N}(x)$, and ${\bf E}_xf_i(X_{t({}^{*}\!\sigma
)})=v_{i,N}(x)$, 
$i=1,2,\ldots ,p$.
\end{theorem}

%%%<--------------

\section{Infinite horizon game\label{infinite}} 

In this class of games the equilibrium strategy is presented in Definition
\ref {equdef} but in class of SS 
\[
\frS_f^{*}=\{\sigma \in \frS^{*}:{\bf E}_xf_i^{-}(X_{t(\sigma )})<\infty
\quad
\mbox{ for every } \ x\in \bbE\mbox{, }i=1,2,\ldots ,p\}. 
\]
Let ${}^{*}\!\sigma \in \frS_f^{*}$  be an equilibrium strategy. Denote
\[
v_i(x)={\bf E}_xf_i(X_{t({}^{*}\!\sigma )}). 
\]

Let us assume that ${}^{(n+1)*}\!\sigma \in \frS_{f,n+1}^{*}$ is constructed and it is an equilibrium strategy. If players $j=1,2,\ldots ,p$, $j\neq i$, apply at moment $n$ the equilibrium strategies ${}^{*}\!\sigma _n^j$ , player $i$ the strategy $\sigma _n^i$ defined by stopping set ${\cal{C}}^{i}$ and ${}^{(n+1)*}\!\sigma $ at moments $n+1,n+2,\ldots$, then the expected payoff of the player $i$, when history of the process up to moment $n-1$ is known, is given by 
\[
  \varphi _n(X_{n-1},{\it{C}}^i)
  ={\bf E}_{X_{n-1}}\left[f_i(X_n) 
  \one_{{}^{i*}\!D_n(D_n^i)}+
  %{\bf E}_x
  v_i(X_n)\one_{\overline{{}^{i*}\!D_n(D_n^i) }}\right], 
\]
where ${}^{i*}\!D_n(A)=\Pi ({}^{*}\!D_n^1,\ldots ,{}^{*}\!D_n^{i-1},A,{}^{*}\!D_n^{i+1},\ldots ,{}^{*}\!D_n^p)$, $ {}^{*}\!D_n^j=\{\omega \in \Omega:{}^{*}\!\sigma _n^j=1\}$, $j=1,2,\ldots ,p$,
$j\neq i$, and $D_n^i=\{\omega \in \Omega :\sigma _n^i=1\}=1\}= \{\omega \in \Omega :X_n\in \cal{C}^i\}$. By Lemma \ref{optimal} the conditional expected gain $\varphi _n(X_{n-1},{\it{C}}^i)$ attains the maximum on the stopping set ${}^{*}\!{\it{C}}_n^i=\{x\in \bbE:f_i(x)\geq v_i(x)\}$ and 
\begin{eqnarray*}
 \varphi _n(X_{n-1},{}^{*}\!{\it{C}}^i)
 &=&{\bf E}_x[(f_i(X_n)-v_i(X_n))^{+}
 \one_{{}^{i*}\!D_n(\Omega )}|\frF_{n-1}] \\
 &&- {\bf E}_x[(f_i(X_n)-v_i(X_n))^{-}
 \one_{{}^{i*}\!D_n(\emptyset )}|\frF_{n-1}] \\
 &&+ {\bf E}_x[v_i(X_n)|\frF_{n-1}].
\end{eqnarray*}

Let us assume that there exists solution $(w_1(x),w_2(x),\ldots ,w_p(x))$
of the equations
\setcounter{equation}{0}
\begin{eqnarray}
w_i(x) &=&
{\bf E}_x(f_i(X_1)-w_i(X_1))^{+}\one_{{}^{i*}\!D_1(\Omega )}
\label{equvalue} \\
&&-{\bf E}_x(f_i(X_1)-w_i(X_1))^{-}\one_{{}^{i*}\!D_1(\emptyset )} 
+ {\bf E}_xw_i(X_1),
\nonumber
\end{eqnarray}
$i=1,2,\ldots ,p$. Consider the stopping game with the following payoff
function for 
$i=1,2,\ldots ,p$.
\[
\phi _{i,N}(x)=\left\{
\begin{array}{ll}
f_i(x) & \mbox{if }n<N, \\
v_i(x) & \mbox{if }n\geq N.
\end{array}
\right.
\]

\begin{lemma}\label{auxinfgame}
Let ${}^{*}\!\sigma {}\in \frS_f^{*}$ be an equilibrium strategy in the infinite horizon game $\cal{G}$. For every $N$ we have 
\[
{\bf E}_x\phi _{i,N}(X_{t^{*}})=v_i(x).
\]
\end{lemma}

Let us assume that for $i=1,2,\ldots ,p$ and every $x\in \bbE$ we have 
\begin{equation}
{\bf E}_x[\sup\nolimits_{n\in \BbbN}f_i^{+}(X_n)]<\infty . \label{supplus} 
\end{equation}

\begin{theorem}
Let $(X_n,\frF_n,{\bf P}_x)_{n=0}^\infty $ be a homogeneous Markov chain and 
the payoff functions of the players fulfill (\ref{supplus}). If $t^{*}=t({}^{*}\!\sigma )$, 
${}^{*}\!\sigma \in \frS_f^{*}$ then ${\bf E}_xf_i(X_{t^{*}})=v_i(x)$.
\end{theorem}

\begin{theorem}
Let the stopping strategy ${}^{*}\!\sigma \in \frS_f^{*}$ be defined by the 
stopping sets
${}^{*}\!{\it{C}}_n^i=\{x\in \bbE:f_i(x)\geq v_i(x)\}$, $i=1,2,\ldots ,p$, 
then ${}^{*}\!\sigma $ is the equilibrium strategy in the infinite stopping
game 
$\cal{G}$.
\end{theorem}

\section{\label{strategiesOFsensors}Determining the strategies of sensors}
Based on the model constructed in Sections~\ref{disorderONsensor}--\ref{noncooperative} for the net of sensors with the fusion center determined by a simple game, one can determine the rational decisions of each nodes. The rationality of such a construction refers to the individual aspiration for the highest sensitivity to detect the disorder without false alarm. The Nash equilibrium fulfills requirement that nobody deviates from the equilibrium strategy because its probability of detection will be smaller. The role of the simple game is to define wining coalitions in such a way that the detection of intrusion to the guarded area is maximal and the probability of false alarm is minimal. The method of constructing the optimum winning coalitions family is not the subject of the research in this article. However, there are some natural methods of solving this problem.    

The research here is focused on constructing the solution of the non-cooperative stopping game as to determine the detection strategy of the sensors. To this end, the game analyzed in Section~\ref{noncooperative} with the payoff function of the players defined by the individual disorder problem formulated in Section~\ref{disorderONsensor} should be derived. 

The proposed model disregards correlation of the signals. It is also assumed that the fusion center has perfect information about signals and the information is available at each node. The further research should help to qualify these real needs of such models and to extend the model to more general cases. In some type of distribution of sensors, e.g. when the distribution of the pollution in the given direction is observed, the multiple disorder model should work better than the game approach. In this case the \emph{a priori} distribution of disorder moment has the form of sequentially dependent random moments and the fusion decision can be formulated as the threshold one: stop when $k^*$ disorder is detected. The method of a cooperative game was used in \cite{ghakri10:cooperativeMR2780188} to find the best coalition of sensors in the problem of the target localization. The approach which is proposed here shows possibility of modelling the detection problem by multiple agents at a general level. 

\section{\label{finalremarks}Final remarks}
In a general case the consideration of this paper leads to the algorithm of constructing the disorder detection system. 

\subsection{Algorithm}
\begin{enumerate}
\item Define a simple game on the sensors.
\item Describe signal processes and \emph{a priori} distribution of the disorder moments at all sensors. Establish the \emph{a posteriori} processes: $\vecPi_n=(\Pi_{1n},\ldots,\Pi_{mn})$, where $\Pi_{kn}=\bP(\theta\leq n|\cF_n)$. 
\item Solve the multivariate stopping game on the simple game to get the individual strategies of the sensors.
\end{enumerate}

%\bibliographystyle{splncs03}
%\bibliography{compet,stgame,PUBLKSz2v2011,DisorderWSKSZ}
\end{document}